\documentclass[a4paper,12pt]{article}
\usepackage{hyperref,amsfonts}

\begin{document}

\title{\bf A generalized Kowalevski Hamiltonian and new integrable 
cases on
 $e(3)$ and $so(4)$. }

\author{Vladimir V.Sokolov \\\\
Landau Institute for Theoretical Physics, \\ Kosygina 2, Moscow 
117334,
Russia
\\\\ email: sokolov@landau.ac.ru
}
\date{}
\maketitle


\vspace{1cm}

\section{A new integrable case for the Kirchhoff equation. }

The Kirchhoff equations (see for example \cite{kozlov}) for the motion 
of a
rigid body in an ideal fluid read as follows
\begin{equation}
M_{t}=M \times \frac{\partial H}{\partial M}+\Gamma \times
\frac{\partial H}{\partial \Gamma}, \qquad \Gamma_{t}=\Gamma
\times \frac{\partial H}{\partial M},
\label{kirhgen}
\end{equation}
where $M=(M_1,M_2,M_3)$ and $\Gamma=(\gamma_1,\gamma_2,\gamma_3)$ are 
three
dimensional vectors, and $\times$ stands for the vector
product. Without loss of generality the Hamiltonian $H$
can be taken in the form
\begin{equation}
H=\sum a_i \, M_i^2+\sum b_{ij}\, (\gamma_i M_j+\gamma_j M_i)+\sum
c_{ij} \, \gamma_i \gamma_j.
\label{GENHAM}
\end{equation}
Apart from $H$, for arbitrary values of parameters $a_i, b_{ij}, 
c_{ij}$
equation (\ref{kirhgen}) has  the following first integrals
\begin{equation}
 I_1=\gamma_1^2+\gamma_2^2+\gamma_3^2,
 \qquad I_2=M_1 \gamma_1+M_2 \gamma_2+M_3 \gamma_3.   \label{kazim}
\end{equation}
These integrals are Casimirs for the Poisson
structure
\begin{equation} \label{puas1}
\{M_{i},M_{j}\}=\varepsilon_{ijk}\,M_{k}, \qquad
\{M_{i},\gamma_{j}\}=\varepsilon_{ijk}\,\gamma_{k}, \qquad
\{\gamma_{i},\gamma_{j}\}=0
\end{equation}
on $e(3)$, which corresponds (see \cite{novik}) to equations 
(\ref{kirhgen}). Therefore for 
integrability of (\ref{kirhgen})
we need one additional first integral 
$I_3$, functionally
independent of $H, I_1, I_2$.

There are  classical integrable cases found by
Kirchhoff, Clebsch and Steklov-Lyapunov \cite{steklov}.
For all these cases the matrices $B=\{b_{ij}\}$ and $C=\{c_{ij}\}$ are
diagonal and the Hamiltonian is of the form
$$
\begin{array}{l}
H=a_{1} M_1^2+a_{2} M_2^2+a_{3} M_3^2+2 b_{11}\,M_1 \gamma_1+
2 b_{22}\,M_2 \gamma_2+
2 b_{33}\,M_3 \gamma_3+ \\[3mm]
\qquad  c_{11} \gamma_1^2+c_{22} \gamma_2^2+c_{33} \gamma_3^2.
\end{array}
$$
The Kirchhoff case is described by the relations
$$
a_1=a_2, \qquad b_{11}=b_{22}, \qquad c_{11}=c_{22}.
$$
For the  Clebsch and Steklov-Lyapunov cases the coefficients $a_i$
are arbitrary and the remaining parameters satisfy the following 
conditions
$$
b_{11}=b_{22}=b_{33},
$$
$$
\frac{c_{11}-c_{22}}{a_3}+\frac{c_{33}-c_{11}}{a_2}+\frac{c_{22}-c_{33
}}{a_1}=0
$$
and
$$
\frac{b_{11}-b_{22}}{a_3}+\frac{b_{33}-b_{11}}{a_2}+\frac{b_{22}-b_{33
}}{a_1}=0,
$$
$$
c_{11}-\frac{(b_{22}-b_{33})^2}{a_1}=
c_{22}-\frac{(b_{33}-b_{11})^2}{a_2}=
c_{33}-\frac{(b_{11}-b_{22})^2}{a_3},
$$
respectively. For each of these three cases there exists an additional
quadratic integral.

In the paper \cite{sokolov} the following Hamiltonian
\begin{equation}
\begin{array}{l}
H=M_{1}^{2}+M_{2}^{2}+2 M_{3}^{2}+ 2 M_{3}\,(c_{1} \gamma_{1}+c_{2} 
\gamma_{2} )
+2 (c_{1} M_{1}+c_{2} M_{2})\, \gamma_{3}+\\[3mm]
\qquad \,\, 4 (c_{2} \gamma_{1}-c_{1} \gamma_{2})^{2}
-4 (c_{1}^{2}+c_{2}^{2}) \, \gamma_{3}^{2}
\label{HAM}
\end{array}
\end{equation}
was considered. It turns out that (\ref{HAM}) is in involution with 
some polynomial
$I_{3}$ of fourth degree with respect to brackets (\ref{puas1}) or, 
equivalently, 
the corresponding Kirchhoff equations (\ref{kirhgen}) possess an
additional fourth degree integral.

Obviously, $I_{3}$  is defined up to a quadratic form of $I_{1}$,
$I_{2}$, and $H.$ In \cite{sokolov} this form has been choosen in such 
a way that
$I_{3}$ depends on a minimal number of new variables.
These variables
\begin{equation} \label{newvar}
\begin{array}{l}
x_1=M_3-c_{1} \gamma_1-c_{2} \gamma_2, \qquad x_2=M_1+c_{1} \gamma_3,
\qquad x_3=M_2+c_{2} \gamma_3, \\[2mm]
x_4=3\,(c_{2} \gamma_1-c_{1} \gamma_2), \qquad x_5=M_3+2 c_{1}
\gamma_1+2 c_{2} \gamma_2, \qquad x_6=\gamma_3
\end{array}
\end{equation}
are very useful for all computations related to the Hamiltonian 
(\ref{HAM}).

In terms of these variables the additional integral for the Kirchhoff 
equations with
the Hamiltonian (\ref{HAM}) has the form
\begin{equation} \label{I3}
I_3=x_1^2\, P+Q^2,
\end{equation}
where
$$
P=(c_1^2+c_2^2)\, x_5^2+(c_2 x_2-c_1 x_3)^2,
\qquad
Q=c_2 x_2 x_4+c_1 x_2 x_5-c_1 x_3 x_4+c_2 x_3
x_5.
$$
It follows from these formulas that $I_3$ does not depend on the 
variable
$x_6.$

Notice that one of the quadratic integrals becomes specially simple in
 
the 
variables (\ref{newvar}):
$$
H+\frac{5}{2}(c_1^2+c_2^2)\, I_1=\frac{1}{2}
(x_1^2+x_2^2+x_3^2+x_4^2+x_5^2).
$$

Simultaneous linear transformations of $M$ and $\Gamma$ with the 
orthogonal matrix of the form
\begin{equation} \label{TT}
T=\pmatrix{\cos{\phi}&\sin{\phi}&0
\cr
-\sin{\phi}&\cos{\phi}&0
\cr
0&0&1
\cr}
\end{equation}
preserve the form (\ref{HAM}) of $H$ while changing the parameters as 
follows
$$
\bar c_{1}=\cos{\phi}\,\, c_{1}-\sin{\phi}\,\, c_{2}, \qquad
\bar c_{2}=\sin{\phi}\,\, c_{1}+\cos{\phi}\,\, c_{2}.
$$
Therefore the only invariant is $\delta=c_{1}^2+c_{2}^2.$

The parameter $\delta$ can be normalized by a rescaling of
$\gamma_i$. If $\delta \ne 0,$ we can take
$c_{1}=1, \, c_{2}=0.$
In addition, there exists a particular (complex) case
$c_{2}=i\, c_{1}.$ Here one can put $c_{1}=1$.

According to a classification theorem (see \cite{sokolov}), the 
Kirchhoff
equation with Hamiltonian (\ref{HAM}) is the only integrable case with
$a_{1}=a_{2}\ne a_{3}$ and an additional first integral of fourth 
degree.

It was noticed by Borisov and Mamaev (see \cite{bms}) that the
change of variables $\bar \gamma_{i}=\gamma_{i}$
\begin{equation} \label{newvari}
\bar M_1=M_1+c_{1} \gamma_3,\qquad \bar M_2=M_2+c_{2} \gamma_3,\qquad
\bar M_3=M_3-c_{1} \gamma_1-c_{2} \gamma_2,
\end{equation}
defined by the first three of the variables (\ref{newvar}),  preserves
brackets (\ref{puas1}) and simplifies the Hamiltonian. Namely 
subtracting
 $4 (c_{1}^{2}+c_{2}^{2})\,I_{1}$ from (\ref{HAM}) and performing
 transformation (\ref{newvari}), we get a new form of the Hamiltonian
\begin{equation} \label{newHAM}
\bar H=M_{1}^2 + M_{2}^2 + 2\, M_{3}^2 + 2\, (a_{1} \gamma_{1} +
a_{2} \gamma_{2})\, M_{3} - (a_{1}^2 + a_{2}^2)\,\gamma_{3}^2,
\end{equation}
where we have put $a_{1}=3 c_{1}, \, a_{2}=3 c_{2}$. In the sequel we 
shall use this
more elegant form of the Hamiltonian.

Borisov and Mamaev \cite{bms} have observed also that the freedom in 
the definition of
$I_{3}$ can be used to bring $I_{3}$  to a factorized form
$I_{3}=k_{1}\, k_{2}$. In terms of the variables (\ref{newvari}) the 
factors are
given by $k_{1}=M_{3}$ and
$$
\begin{array}{l}
k_{2}=(M_{1}^2 + M_{2}^2+ M_{2}^2)\, M_{3} +
2\,(a_{1} M_{1} + a_{2} M_{2})\,(M_{1} \gamma_{1} + M_{2} 
\gamma_{2})\\[3mm]
\qquad +2\,(a_{1}\gamma_{1}+a_{2}\gamma_{2})\,M_{3}^{2}
    +(a_{1}\gamma_{1}+a_{2}\gamma_{2})^{2}\,M_{3} \\[3mm] \qquad -
    (a_{1}^2+a_{2}^2)\,(2 M_{1} \gamma_{1}+2 
M_{2}\gamma_{2}+M_{3}\gamma_{3})
    \,\gamma_{3}
.
\end{array}
$$
It turns out that both $k_{1}=0$ and $k_{2}=0$ are invariant 
submanifolds. This fact 
follows from the following formulas:
$$
\dot k_{1}=2 \, (a_{2} \gamma_{1}-a_{1}\gamma_{2})\,k_{1}, \qquad
\dot k_{2}=-2 \, (a_{2} \gamma_{1}-a_{1}\gamma_{2})\, k_{2}.
$$

The particular case $a_{1}^{2}+a_{2}^{2}=0$ is simpler than the 
generic
case. For this complex case a Lax representation has been found in
\cite{sokolov}. For the case $a_1^2+a_2^2\ne 0$ Lax representations 
have not been 
found yet.

\section{The corresponding integrable case on $so(4)$.}

Let us consider the standard deformation
\begin{equation} \label{puas2}
\{M_{i},M_{j}\}=\varepsilon_{ijk}\,M_{k}, \qquad
\{M_{i},\gamma_{j}\}=\varepsilon_{ijk}\,\gamma_{k}, \qquad
\{\gamma_{i},\gamma_{j}\}=x\, \varepsilon_{ijk}\,M_{k}.
\end{equation}
of bracket (\ref{puas1}), where
$x$ is an arbitrary parameter. It is well known that if $x>0$ then
(\ref{puas2}) defines a Poisson bracket on  $so(4)$. The Casimirs of 
this
bracket are given by
\begin{equation}
\label{sokol5}
I_1=x\,(M_{1}^{2}+M_{2}^{2}+M_{3}^{2})+
(\gamma_{1}^{2}+\gamma_{2}^{2}+\gamma_{3}^{2}),\qquad
I_2=M_{1} \gamma_{1}+M_{2} \gamma_{2}+M_{3} \gamma_{3}.
\end{equation}

The main result of the paper \cite{bms} consists in the following 
observation.
The Hamiltonian (\ref{newHAM}) possesses a commuting integral of 
fourth degree
$\bar I_{3}=\bar k_{1}\, \bar k_{2}$, where
$$
\bar k_{1}=k_{1}, \qquad \bar k_{2}=k_{2}+x\, (a_{2} M_{1} - a_{1} 
M_{2})^2\,
M_{3}.
$$
This integrable deformation of the integrable case from Section 1 has 
been
found independently by Sokolov and Borisov-Mamaev. Here we present the 
most
elegant form of the deformed additional integral proposed by Borisov 
and Mamaev.

Note that the dynamical system with Hamiltonian (\ref{newHAM}) and 
brackets
(\ref{puas2}) has just two one-parameter families of solutions of the 
form
$M_{i}=X_i\,t^{-1}, \,\gamma_{i}=Y_i\,t^{-1}$. First of them is 
defined by
$$
X_{3}=Y_{3}=0, \quad Y_{1}^{2}+Y_{2}^{2}=0,
\quad 2 a_{1} Y_{2}- 2 a_{2} Y_{1}=1,
\quad X_{1}=2 X_{2}\,(a_{1} Y_{1}+a_{2} Y_{2}).
$$
For the second family we have
$$
X_{1}=a_{1} Y_{3}, \qquad X_{2}=a_{2} Y_{3}, \qquad
X_{3}=-a_{1} Y_{1}-a_{2} Y_{2},
$$
$$
\Big(1+x\,(a_{1}^{2}+a_{2}^{2})\Big)\,
\Big(Y_{1}^{2}+Y_{2}^{2}+Y_{3}^{2}\Big)=\frac{x}{4}, \qquad
2 a_{1} Y_{2}- 2 a_{2} Y_{1}=-1.
$$
For both families the Kowalewski exponents are $\{-1,0,1,2,2,2\}$.

\section{Generalizations. }

This Section contains new results concerning possible integrable 
generalizations 
of the Hamiltonian (\ref{newHAM}). 

It turns out that extra linear terms can be added to the Hamiltonian 
(\ref{newHAM}). 
The most general form of such terms is given by 
\begin{equation} \label{tailHAM}
\begin{array}{l}
H=M_{1}^2 + M_{2}^2 + 2\, M_{3}^2 + 2\, (a_{1} \gamma_{1} +
a_{2} \gamma_{2})\, M_{3} - (a_{1}^2 + a_{2}^2)\,\gamma_{3}^2+\\[3mm]
 \qquad k_1 (M_3+a_1 \gamma_1+a_2 \gamma_2)+k_2 (a_1 \gamma_2-a_2 
\gamma_1)
,
\end{array}
\end{equation}
where $k_i$ are arbitrary constants. 
This Hamiltonian possesses an additional first integral of fourth 
degree on $so(4)$. 
Historically, the author found first that this is true for Hamiltonian 
(\ref{tailHAM}) on $e(3)$  
with $k_1=0.$ After that, independently, V.Kuznetsov and A.Tsiganov 
have informed the author that the same 
Hamiltonian is integrable on $so(4)$. Finally, the term with non-zero 
$k_1$ was added by the author very 
recently.

Since the explicit expression for the additional integral $I_4$ is 
rather long, we put $a_1=0$ in the 
Hamiltonian (\ref{tailHAM}) and rewrite it as follows 
\begin{equation} \label{tailredHAM}
\tilde H=M_{1}^2 + M_{2}^2 + 2\, M_{3}^2 + 2\, \lambda_{1} \gamma_{2} 
M_{3} - \lambda_{1}^2 \,\gamma_{3}^2+
2 \lambda_2 (M_3+\lambda_1 \gamma_2)+2 \lambda_3 \gamma_1,   
\end{equation}
where $\lambda_i$ are arbitrary constants. If $\lambda_1=\lambda_2=0,$ 
then the Hamiltonian 
reduces just to the famous Kowalewski Hamiltonian. The case 
$\lambda_1=0$ corresponds to the Kowalewski Hamiltonian with the 
additional gyrostatic term. 

For the shortened Hamiltonian (\ref{tailredHAM}) the additional 
integral has the form 
$$
\begin{array}{l}
          I_4=M_3^2(M_1^2 + M_2^2 + M_3^2)+
          
            \lambda_1 M_3\Big(2 M_1 M_2 \gamma_1 +2 M_2^2 \gamma_2+ 2  
M_3^2 \gamma_2 +  \\[3mm]
            
                  \qquad \lambda_1 M_3 \gamma_2^2 - 2 \lambda_1 M_1 
\gamma_1 \gamma_3 - 2 \lambda_1 
                  M_2 \gamma_2 \gamma_3 -
                  \lambda_1 M_3 \gamma_3^2\Big)+\\  
            
            \qquad 2 \lambda_2
              \Big(M_3(M_1^2+ M_2^2 + M_3^2) + \lambda_1 M_1 M_2 
\gamma_1 +  
                  \lambda_1 M_2^2 \gamma_2 + 2 \lambda_1 M_3^2 
\gamma_2 + \\[3mm]
                  \qquad \lambda_1^2 M_3 \gamma_2^2 - 
\lambda_1 M_2 
M_3 \gamma_3 -
                  \lambda_1^2 M_1 \gamma_1 \gamma_3 - \lambda_1^2 M_2 
\gamma_2 \gamma_3\Big)+\\[3mm]
                   
            \qquad 2 \lambda_3 
              \Big(M_1^2 \gamma_1 + M_3^2 \gamma_1 + M_1 M_2 \gamma_2 
+ \lambda_1 M_3 \gamma_1 \gamma_2\Big)-
\\[3mm]
            
           \qquad \lambda_2^2 M_3^2+2 \lambda_2 \lambda_3 M_3 
\gamma_1+2 \lambda_1 \lambda_2 \lambda_3 \gamma_1 \gamma_2
            +\lambda_1^2 \lambda_2^2 \gamma_2^2-\lambda_3^2 
\gamma_2^2-\\[3mm]
            \qquad 2 \lambda_2 \lambda_3 M_1 \gamma_3-2 \lambda_1 
\lambda_2^2 M_2 \gamma_3+
            2 \lambda_1^2 \lambda_2^2 \gamma_3^2
            
            - 2 \lambda_2^2 \Big(\lambda_2 M_3 + \lambda_3 \gamma_1 
+\lambda_1 \lambda_2 \gamma_2\Big) +\\[3mm]

             \qquad x \Big(\lambda_1^2 M_1^2 M_3^2+2 \lambda_1^2 
\lambda_2 M_1^2 M_3 - 
             2 \lambda_1 \lambda_3 M_1 M_2 M_3 +\\[3mm]
             \qquad \lambda_1^2 \lambda_2^2 M_1^2 - \lambda_3^2 M_1^2 
- 2 \lambda_1 \lambda_2 \lambda_3 M_1 M_2 
                  - \lambda_3^2 M_3^2 
                  \Big)
.
\end{array}
$$
To get the integral $I_4$ on $e(3)$ we can simply put $x=0$ in this 
formula. 
The explicit form of the integral $I_4$ on $so(4)$ for (\ref{tailHAM}) 
can be easily reconstructed with 
the help of transformation (\ref{TT}).

For the next generalization let us consider the the following 
Hamiltonian 
\begin{equation} \label{genHAM}
\tilde H=M_{1}^2 + M_{2}^2 + 2\, M_{3}^2 + 2\, (a_{1} \gamma_{1} +
a_{2} \gamma_{2})\, M_{3} - (a_{1}^2 + a_{2}^2)\,\gamma_{3}^2+
F(\gamma_{1},
\gamma_{2}). 
\end{equation}
on $e(3).$ It is not difficult to verify that the corresponding 
Kirchhoff equations 
possess an additional integral of fourth degree of the form 
$\bar I_3=k_1 k_2+T$, where $T$ is a cubic polynomial in $M_1,M_2,M_3$ 
with 
coefficients dependent on $\gamma_{1},
\gamma_{2}$ iff 
$$
\displaystyle F=2 \lambda_1 \, (a_2 \gamma_1-a_1 \gamma_2)+
\frac{\lambda_2 }{\sqrt{\gamma_{1}^2+
\gamma_{2}^2}},
$$
where $\lambda_i$ are arbitrary constants.
In this case the function $T$ reads as follows
$$
\begin{array}{l}
T=2 a_2 \lambda_1 \gamma_1 \, M_1^2 + 2 \lambda_1 (a_2 \gamma_2 -  a_1 
  \gamma_1)\, M_1 M_2 -
  2 a_1 \lambda_1 \gamma_2 \, M_2^2 \ + \\[3mm]
  \qquad \displaystyle \left(2 a_2 \lambda_1 \gamma_1-2 a_1 \lambda_1 
\gamma_2+ \frac{\lambda_2 }{\sqrt{\gamma_{1}^2
  +
\gamma_{2}^2}}\right) \, M_3^2+ 
  \\[3mm]
  \qquad \displaystyle  (a_1 \gamma_1+ a_2 \gamma_2)
  \left(2 a_2 \lambda_1 \gamma_1-2 a_1 \lambda_1 \gamma_2+  
\frac{\lambda_2}{\sqrt{\gamma_{1}^2+
\gamma_{2}^2}}\right) 
  \, M_3-\\[3mm]
  \qquad \displaystyle  \lambda_1 \left(\lambda_1 (a_1 \gamma_1+a_2 
\gamma_2)^2+(a_1 \gamma_2-a_2 \gamma_1) 
   \frac{\lambda_2}{\sqrt{\gamma_{1}^2+
\gamma_{2}^2}}\right).
\end{array}
$$

Performing the scaling $\gamma_i \rightarrow \varepsilon \gamma_i$ and 
redefining the constants $\lambda_i,$ we 
get the following integrable Hamiltonian
\begin{equation} \label{genKow}
\begin{array}{l}
\tilde H=M_{1}^2 + M_{2}^2 + 2\, M_{3}^2 + 2\varepsilon \, (a_{1} 
\gamma_{1} +
a_{2} \gamma_{2})\, M_{3} - \varepsilon^2 (a_{1}^2 + 
a_{2}^2)\,\gamma_{3}^2+\\[3mm]
\qquad \displaystyle 2 \lambda_1 \, (a_2 \gamma_1-a_1 
\gamma_2)+\frac{\lambda_2 }{\sqrt{\gamma_{1}^2+
\gamma_{2}^2}}.
\end{array}
\end{equation}
The case $\varepsilon=0$ corresponds to a known generalization of the 
Kowalewski Hamiltonian found by H.M. Vehia 
\cite{vehia}. 

Very recently A. Tsiganov and the author have found (see 
\cite{tsigan}) the following 
generalization of the Goryachev-Chaplygin Hamiltonian:
\begin{equation} \label{genGOR}
\begin{array}{l}
H= M_{1}^2 + M_{2}^2 + 4\, M_{3}^2 + 2 \lambda_1 \gamma_1+2 \lambda_2 
\gamma_2+\lambda_3 M_3+\\[3mm]
\qquad 4 (a_1 \gamma_1+a_2 \gamma_2)\, M_3-(a_1^2+a_2^2)\gamma_3^2.
\end{array}
\end{equation}
On the fixed level 
$I_2=0$ of the Casimir $I_2$ the Hamiltonian $H$ commutes with 
$$
\begin{array}{l}
I_3=(M_1^2+M_2^2)\,M_3-2 (a_1 M_1+a_2 M_2)\, M_3 
\gamma_3+(a_1^2+a_2^2)\,M_3 \gamma_3^2 \\[3mm]
\qquad -4 \lambda_3 M_3^2-4 \lambda_3 (a_1 \gamma_1+a_2 \gamma_2)\, 
M_3-(\lambda_1 M_1+\lambda_2 M_2)\,\gamma_3 \\[3mm]
\qquad +(a_1 \lambda_1+a_2 \lambda_2)\, \gamma_3^2-4 \lambda_3^2 M_3-2 
\lambda_3 (\lambda_1 \gamma_1+\lambda_2 \gamma_2).
\end{array}
$$
If $a_1=a_2=\lambda_3=0,$ then (\ref{genGOR}) coincides with the 
Goryachev-Chaplygin Hamiltonian. The integrable case 
$a_1=a_2=0$, $\lambda_3\ne 0$ was found in (\cite{komkuz}).

\section{Historical remarks.}

A fourth degree integral for the Kirchhoff equation on $e(3)$ had been 
found
by S.\,A.\, Chaplygin under the additional assumption that
 the scalar 
product 
$(M,\Gamma)$ is equal to zero ~\cite{chaplyg14,kuztsi}.
The corresponding partially integrable case  on $so(4)$ was found by
O.\,I.\,Bogoyav\-lenski ~\cite{bogo15}.

An integrable case on $so(4)$ with an integral of fourth degree was 
found
by M.\,Adler and P.\,van~Moerbeke~\cite{adler2}. A Lax pair for that 
case
has been constructed by A.\,Reyman and M.\,Semenov-Tian 
-Shansky~\cite{reysem}.
The fact that $k_{1}=0$ is the invariant relation has been used in
\cite{sokolov} to prove that the Kirchhoff equation with Hamiltonian
(\ref{HAM}) is not linearly equivalent to the Adler-van Moerbeke-
Reyman-Semenov-tian -Shansky case on $so(4)$.

R.\, Liouville~\cite{liouv1} has obtained necessary conditions for the 
existence
of algebraic integrals for the Kirchhoff equation with non-diagonal 
matrix $B$.

In recent investigations of algebraic integrability~\cite{sadetov3} it 
was
assumed that all matrices~$\bf A, B, C$ are diagonal. 
In~\cite{kozlov9,borisov2}
the matrix~$\bf A$ was assumed to be defined by the 
inertial tensor~$\bf I$ of a real rigid body:~${\bf A}={\bf I}^{-1}$, 
and all
entries $a_{i}$ to be distinct. The latter restriction has been also 
imposed in the
papers \cite{vesel,adler}, devoted to integrability conditions of the 
equations
on $so(4)$.

{\bf Acknowledgments.} The author is grateful to T. Wolf for his help 
during the
work with the package "Crack" \, and to A.V. Borisov, Yu. N. Fedorov, 
V. B. Kuznetsov,
I.S. Mamaev, and A. V. Tsiganov for useful 
discussions. The
research was partially supported by RFBR grant 99-01-00294,
INTAS grant 99-1782, and EPSRC grant GR K99015.

====


\begin{thebibliography}{99}

\bibitem{kozlov} Kozlov V. V. Symmetries, topology and resonances in 
Hamiltonian
mechanics, {\em Publ. Udmurdski state university:
Izhevsk}, 1995.

\bibitem{novik} Novikov S. P. Hamiltonian formalism and multi-valued 
analog of
Morse theory, {\em UMN}, {\bf 37}(5), 3--49, 1982.

\bibitem{steklov} Steklov V. A. On the motion of a rigid body in a 
fluid,
{\em Kharkov}, 234 pages,  1893.

\bibitem{sokolov} Sokolov V. V. A new integrable case for the 
Kirchhoff equation,
{\em Theor. and Math. Physics}, {\bf }(129)(1), 31--37, 2001.

\bibitem{bms} Borisov A. V., Mamaev S. I., and Sokolov V. V.
A new integrable case on $so(4))$, {\em Doklady RAN}, {\bf }(), , 
2001.


\bibitem{vehia} Vehia H. M. 
New generalizations of the integrable problems in rigid body dynamics, 
{\em J. Phys. A}, {\bf 30}, 7269--7275, 1997.

\bibitem{tsigan} Sokolov V. V., Tsiganov A. V.
On integrable deformation of Goryachev-Chaplygin top, {\em submitted 
to Theor. and Math. Physic}, 
{\bf }, , 2002.

\bibitem{komkuz} Komarov I. V., Kuznetsov V. B. 
Genaralized Goryachev-Chaplygin gyrostat in quantum mechanics, 
{\em Zapiski seminarov LOMI}, {\bf IX}, 134--141, 1987.


\bibitem{chaplyg14} Chaplygin S.\,A.
A new partial solution of the problem of rotation of heavy rigid body 
around
a fixed point, {\em Sobranie sochineny}, v.~1, M.-L.: GITTL, 1948, 
p.~125--132.

\bibitem{kuztsi} Kuznetsov V. B., Tsiganov A. V.  
A special case of Neumann's system and the 
Kowalewski-Chaplygin-Goryachev top, 
{\em J. Phys. A}, {\bf 22}, L73--L79, 1989.

\bibitem{bogo15} Bogoyavlenski O.\,I.,
Breaking solitons, nonlinear integrable equations,
{\em M.: Nauka}, 1991, 320~p.


\bibitem{adler2} Adler M., van Moerbeke P.
A new geodesic flow on ${\rm SO}(4)$. Probability,
statistical mechanics, and number theory, {\em Adv. Math.
Suppl. Stud.}, {\bf 9}, 81--96,1986.

\bibitem{reysem} Reyman A. G., Semenov-Tian-Shansky M. A.
A new integrable case of the motion of the 4-dimensional
rigid body, {\em Comm. in Math. Phys.}, {\bf 105},
461--472, 1986.

\bibitem{liouv1} Liouville R.
Sur le mouvement d'un solide dans un liquide ind\'efini, {\em  Comp. 
Rend.
Ac. Sc.}, {\bf ser.~2}, 874--876,  1896.

\bibitem{sadetov3} Sadetov S.\,T.,
Intagrability conditions for the Kirchhoff equation, {\em Vestnik MGU, 
ser. math.
mech.}, {\bf 3}, 56--62, 1990.

\bibitem{kozlov9} Kozlov V.\,V., Onischenko D.\,A.,
On non-integrability of the Kirchhoff equations, {\em DAN SSSR},  {\bf 
226}(6),
1298--1300, 1982.

\bibitem{borisov2} Borisov A.\,V., Necessary and sufficient 
integrability
conditions for the Kirchhoff equations,
{\em Reg. \& Chaot. Dyn.},  {\bf 1}(2), 61--73, 1996.

\bibitem{adler} Adler M., van Moerbeke P. The Kowalevski and 
Henon-Heiles
motions as Manakov geodesic flows on so(4) - a two
dimensional family of Lax pairs, {\em Comm. in Math.
Phys.}, {\bf 113}, 659--700, 1988.

\bibitem{vesel} Veselov A. P. On integrability conditions for the 
Euler equation
on so(4), {\em DAN SSSR}, {\bf 270}(6), 1298--1300, 1983.


\end{thebibliography}
\end{document}